\documentclass[10pt,twocolumn,letterpaper]{article}

\usepackage{wacv}
\usepackage{times}
\usepackage{epsfig}
\usepackage{graphicx}
\usepackage{amsmath}
\usepackage{amssymb}
\usepackage[ruled,vlined,linesnumbered]{algorithm2e}
\usepackage{booktabs}
\usepackage[caption=false, font=footnotesize]{subfig}
\usepackage{pgfplots}
\pgfplotsset{compat=1.14}
\usepackage{comment}
\usepackage{array, makecell} 
\usepackage[numbers,sort,compress]{natbib}
\usepackage{times,epsfig,graphicx,amsmath,amssymb}
\usepackage{times,graphicx,amsmath,amssymb,booktabs,tabulary,multirow,overpic,bbm,amssymb}

\def\wacvPaperID{330} 

\wacvfinalcopy 

\ifwacvfinal
\def\assignedStartPage{1} 
\fi

\ifwacvfinal
\usepackage[breaklinks=true,bookmarks=false]{hyperref}
\else
\usepackage[pagebackref=true,breaklinks=true,colorlinks,bookmarks=false]{hyperref}
\fi

\ifwacvfinal
\else
\fi

\usepackage{amsmath,amsfonts,bm}

\def\eqref#1{equation~\ref{#1}}

\def\1{\bm{1}}

\DeclareMathAlphabet{\mathsfit}{\encodingdefault}{\sfdefault}{m}{sl}
\SetMathAlphabet{\mathsfit}{bold}{\encodingdefault}{\sfdefault}{bx}{n}

\title{Segmentation of Pulmonary Opacification in Chest CT Scans of COVID-19 Patients}

\author{%
    Keegan Lensink\thanks{The University of British Columbia}\textsuperscript{\space\space}\thanks{Xtract AI}%
    \and Issam Laradji\thanks{Element AI}%
    \and Marco Law\footnotemark[1]\textsuperscript{\space\space}\thanks{SapienML}%
    \and Paolo Emilio Barbano\thanks{AWS, Worldwide Public Sector}%
    \and Savvas Nicolaou\footnotemark[4]\textsuperscript{\space\space}\thanks{Department of Radiology, The University of British Columbia}%
    \and William Parker\footnotemark[4]\textsuperscript{\space\space}\footnotemark[6]%
    \and Eldad Haber\footnotemark[1]\textsuperscript{\space\space}\footnotemark[2]
}

\begin{document}
\maketitle

\begin{abstract}
The Severe Acute Respiratory Syndrome Coronavirus 2 (SARS-CoV-2) has rapidly spread into a global pandemic. A form of pneumonia, presenting as opacities with in a patient's lungs, is the most common presentation associated with this virus, and great attention has gone into how these changes relate to patient morbidity and mortality.
In this work we provide open source models for the segmentation of patterns of pulmonary opacification on chest Computed Tomography (CT) scans which have been correlated with various stages and severities of infection.
We have collected 663 chest CT scans of COVID-19 patients from healthcare centers around the world, and created pixel wise segmentation labels for nearly 25,000 slices that segment 6 different patterns of pulmonary opacification.
We provide open source implementations and pre-trained weights for multiple segmentation models trained on our dataset.
Our best model achieves an opacity Intersection-Over-Union score of 0.76 on our test set, demonstrates successful domain adaptation, and predicts the volume of opacification within 1.7\% of expert radiologists.
Additionally, we present an analysis of the inter-observer variability inherent to this task, and propose methods for appropriate probabilistic approaches.
\end{abstract}

The Severe Acute Respiratory Syndrome Coronavirus 2 (SARS-CoV-2 or COVID-19) has rapidly spread into a global pandemic and resulted in over 450,000 COVID-19 related deaths as of June 15th 2020 \cite{whoCovid19}.
While the disease can symptomatically present itself in many ways, ranging from asymptomatic to flu-like symptoms to even a life threatening Acute Respiratory Distress Syndrome (ARDS), the primary and most common presentation relating to all symptoms, is the physical presence of opacities and consolidation in a patient's lungs.
Healthcare centers around the world are overwhelmed and facing shortages of the essential equipment necessary to manage the symptoms of this disease. 
Rapid screening is necessary to diagnose the disease and slow the spread, and effective tools are essential for prognostication in order to efficiently allocate resources to those who need them most.


\begin{figure}
\centering
    \subfloat[\label{fig:op_a}]{%
        \includegraphics[width=0.45\linewidth]{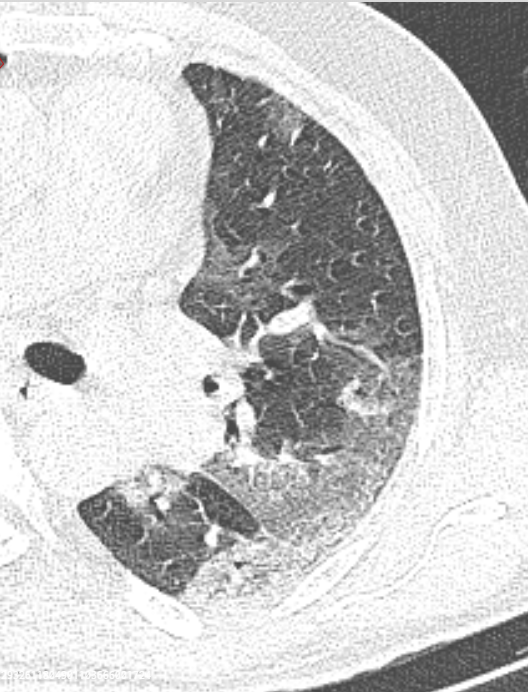}
    }
    \hfill
    \subfloat[\label{fig:op_b}]{%
        \includegraphics[width=0.45\linewidth]{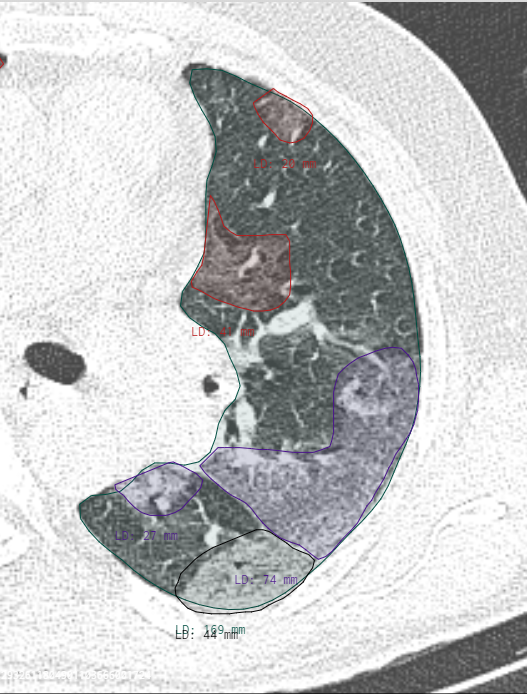}
    }
\caption{A visualization of an axial slice of a CT scan cropped of the left lung. Figure \ref{fig:op_a} shows pulmonary opacification present in a patient with COVID-19, while Figure \ref{fig:op_b} shows the corresponding annotations provided by a radiologist (Red: Pure GGO, Purple: GGO w/ intra- \& inter-lobular Lines (Crazy Paving), and Black: Consolidation).}
\label{fig:op}
\end{figure}

While RT-PCR has emerged as the standard screening protocol for COVID-19 in many countries, the test has been shown to have high false negative rates due to its relatively low sensitivity yet high specificity \cite{zu2020coronavirus}. Recent work has shown that the analysis of chest CT scans by trained radiologists increases the diagnostic sensitivity \cite{ai2020correlation}. 
This is because the virus attacks and inhibits the alveoli of the lung, which fill with fluid in response, causing various forms of opacification within the lung when seen on Computed Tomography (CT) scans. 
Due to an increase in density, these areas present on CT scans as increased attenuation with preserved bronchial and vascular markings, known as a ground glass opacity (GGO).
In addition to this, when the accumulation of fluid progresses to obscure bronchial and vascular regions on CT scans, it is known as consolidation.

In addition to providing complimentary diagnostic properties, the analysis of CT scans has great potential value for the prognostication of patients with COVID-19. 
The percentage of well-aerated-lung (WAL) has emerged as a predictive metric for determining prognosis of patients confirmed with COVID-19, including admission to the ICU and death \cite{Colombi2020WellaeratedLO}.
The quantification of percentage of WAL is often done by visually estimating volume of opacification relative to healthy lung, which is a time consuming process, or can be roughly estimated automatically through attenuation values within the lung.
In addition to the percent of WAL, which does not account for the various forms of opacification, expert interpretation of CT scans can provide insight on the severity of the infection by identifying various patterns of opacification (see Table \ref{tab:classes}). 
The prevalence of these patterns, and the severity of the infection, can also correlate to different stages of the disease \cite{Li2020, Wang2020}. 
Therefore, automatic quantification of both the percentage WAL and the opacification composition could enable efficient estimation of the stage of the disease, and even a glimpse at the risk for poor outcomes.

Standard diagnostics require experienced radiologists, and is highly time consuming. Thus, there is a need to develop machine learning techniques to deal with the problem in a quantitative way to support the radiological team.
The task is difficult for a number of reasons, and there is a limited amount of available work focused on this particular problem.
First, privacy restrictions and labelling cost lead to a lack of available public data.
As a result, there is only one small public dataset \cite{publicTestSet} known to us at the time of writing this manuscript.
For this reason, we are unable to publicly release the collected dataset, at this time.
Second, as we show in this work, the segmentation process has a subjective component. 
Indeed, the collected data is heavily influenced by the instructions provided to the annotators and therefore, it is likely problematic to combine data sets collected under different labelling regimes.  
Third, the cost of acquiring pixel level segmentation labels is prohibitive. 
Compared to common computer vision tasks such as the annotation of street scenes, this task is non-trivial and requires careful attention by highly skilled expert annotators. 
In this study, we found that pixel level annotation required an average of 60 minutes per scan, leading to a total of roughly 660 highly trained person-hours to acquire the dataset. 

From a machine learning point of view, the second point deserves special attention. As we show next,
the segmentation task is especially difficult due to the subjectivity of the labels and the low inter-class variance.
The regions of opacification are, by definition, hazy and without clearly defined borders. 
As such, there is an expected increased level of inter-observer variability. 
In addition, the patterns that we wish to differentiate have low inter-class variability. 
While the patterns described in \ref{tab:classes} are technically independent, they are not mutually exclusive in a given area, and distinguishing these complex patterns is very time consuming and difficult for even expert radiologists. 

\begin{table*}[t]
\small
\begin{center}
\caption{Summary of Classes.}
\begin{tabular}{lrrrr}
\hline
\textbf{Pattern Description}                 & \textbf{Class Number} & \textbf{Group Number} & \textbf{Group Colour} & \textbf{Prevalence (\%)}\\
\hline
Unlabelled                                  & -1                     & -1  & Blue &  3.12350               \\
Background                                   & 0                     & 0  & Orange &  94.70900               \\
Left Lung                                    & 1                     & 1  & Green  & 0.48229               \\
Right Lung                                   & 2                     & 1  & Green &  0.52713               \\
Pleural Effusion                             & 3                     & 0  & Orange & 0.02021               \\
Lymphadenopathy                              & 4                     & 0  & Orange & 0.00002               \\
Pure Ground Glass Opacification              & 5                     & 2  & Red   & 0.33230              \\
GGO w/ Smooth Interlobular Septal Thickening & 6                     & 3  & Purple  & 0.04404               \\
GGO w/ Intralobular Lines (Crazy Paving)     & 7                     & 3  & Purple&  0.44893                \\
Organizing Pneumonia Pattern                 & 8                     & 4  & Brown &  0.12062               \\
GGO w/ Peripheral Consolidation (Atoll Sign) & 9                     & 4  & Brown & 0.01665                \\
Consolidation                                & 10                    & 4  & Brown &  0.17524     
\end{tabular}

\label{tab:classes}
\end{center}
\end{table*}

Given the challenges mentioned above, the goal of this work is to provide open source models for the segmentation of patterns of pulmonary opacification, which have been correlated with various stages and severity of COVID-19 pneumonia. 
While the development of open source models have been hindered by the lack of publicly available data, we hope that by releasing our open source model and pre-trained weights, we can enable healthcare centers and researchers around the world to develop tools for the effective diagnosis and prognostication of COVID-19 on CT scans. 
In addition to our models, we hope to enable researchers by discussing the insights gained from our work into this difficult task, particularly related to the incorporation of uncertainty and the high inter-observer variability between annotators. 
We provide an \href{https://github.com/UBC-CIC/COVID19-L3-Net}{open source software implementation}\footnote{https://github.com/UBC-CIC/COVID19-L3-Net}, with training procedure using a small public dataset, and an \href{https://cic.ubc.ca/projects/open-source-ai-model-for-covid-19-ct-diagnostics-and-prognosis/covid-19-ai-tool-demo/}{online visualization tool}\footnote{https://cic.ubc.ca/projects/open-source-ai-model-for-covid-19-ct-diagnostics-and-prognosis/covid-19-ai-tool-demo/} to easily view predictions on our private dataset. 

Our contributions are as follows:
\begin{enumerate}
    \item We have collected 663 chest CT scans of patients with COVID-19 pneumonia from healthcare centers around the world, and created pixel wise segmentation labels for nearly 25,000 slices that segment 6 different forms of pulmonary opacification that have been correlated with stages and severity of COVID-19.
    \item We provide open source implementations and pre-trained weights for multiple segmentation models trained on our dataset, and show that these models adapt to domains that are withheld from the training set. We hope that by making this publicly available we will ease the burden of development on healthcare centers around the world that have limited access to data.
    \item We present an analysis of the inter-observer variability inherent in this task, and propose solutions through annotation procedure and noise modelling.
\end{enumerate}

\section{Related Work}
\label{sec:related_work}
In this section we discuss the work that is most relevant for this paper. We start with semantic segmentation for CT Scans on general medical problems, followed by semantic segmentation for COVID-19.

Deep learning-based methods have been widely applied in medical image analysis to combat COVID-19~\cite{keshani2013lung, hesamian2019deep, wang2017central}. They have been proposed to detect patients infected with COVID-19 via radiological imaging. For example, a COVID-Net~\cite{wang2020covid} was proposed to detect COVID-19 cases from chest radiography or X-Ray images. An anomaly detection model~\cite{schlegl2017unsupervised} was designed to assist radiologists in analyzing the vast amounts of chest X-ray images. For CT imaging, a location-attention oriented model was employed to calculate the infection probability of COVID-19. A weakly-supervised deep learning-based software system was developed in~\cite{zheng2020deep} using 3D CT volumes to detect COVID-19. Although plenty of AI systems have been proposed to provide assistance in COVID-19 diagnostics for clinical practice, there are only a few related works~\cite{fan2020inf}, and no significant impact has been shown using AI to improve clinical outcomes, as of yet.

\paragraph{Semantic segmentation for CT Scans} has been widely used for diagnosing lung diseases. Diagnosis is often based on segmenting different organs and lesions from chest CT slices, which can provide essential information for doctors to identify underlying disease processes. Many methods exist that perform lung nodule segmentation. Early algorithms are based on SVM to extract features and detect nodule segmentations \cite{keshani2013lung}. Later, algorithms based on deep learning emerged~\cite{hesamian2019deep}. Some examples are central focus CNNs~\cite{wang2017central} and GAN-synthesized data to improve the training of a discriminative model for pathological lung segmentation. Latest methods are based on two deep networks to segment lung tumors from CT slices by adding multiple residual streams of varying resolutions, and multi-task learning of joint classification and segmentation. 

\paragraph{Semantic segmentation for COVID-19} 
While COVID-19 is a recent phenomenon, several methods have been proposed to analyze infected regions of COVID-19 in the lungs. \cite{fan2020inf} proposed a semi-supervised learning algorithm for automatic COVID-19 lung infection segmentation from CT scans. Their algorithm leverages attention to enhance representations. Similarly, \cite{zhou2020automatic} proposed to use spatial and channel attention to enhance representations, and  \cite{chen2020residual} augment UNet~\cite{Ronneberger2015} with ResNeXt~\cite{xie2017aggregated} blocks and attention to improve its efficacy. Instead of focusing on the architecture. Although previous methods are accurate, their computational cost can be prohibitive.
\cite{chaganti2020quantification} similarly propose the segmentation of tomographic patterns from chest CT using a large annotated dataset, and compute measures of the severity of infection. 

\section{Data Collection and Annotation}
\label{sec:dataset}

\begin{table*}[t]
\small
\begin{center}
\caption{Dataset Composition - Slices(Scans).}
\begin{tabular}{lrrrr}
\hline
  \textbf{Region} &
  \textbf{Testing} &
  \textbf{Validating} &
  \textbf{Training} &
  \textbf{All Splits} \\ \hline
Iran         & 1852 (52) & 2645 (95)  & \multicolumn{1}{r|}{12365 (405)} & 16862 (552) \\
Italy        & 75 (1)    & 0 (0)      & \multicolumn{1}{r|}{0 (0)}       & 75 (1)      \\
South Korea  & 1568 (18) & 1151 (20)  & \multicolumn{1}{r|}{3255 (51)}   & 5974 (89)   \\
Vancouver    & 1407 (13) & 0 (0)      & \multicolumn{1}{r|}{0 (0)}       & 1407 (13)   \\
Saudi Arabia & 0 (0)     & 69 (1)     & \multicolumn{1}{r|}{589 (7) }    & 658 (8)     \\ \hline
All Regions  & 4902 (84) & 3865 (116) & \multicolumn{1}{r|}{16208 (463)} & 24975 (663)
\end{tabular}

\label{tab:split}
\end{center}
\end{table*}


We have worked with health centers around the world to retrospectively collect 663 CT scans of patients suspected of having COVID-19. 
The dataset is composed of scans from health centers in Canada, Italy, South Korea, Iran, and Saudi Arabia, where each respective health center's research ethics board approved use of the dataset.
The dataset represents a relatively equal distribution of sex, with 321 female scans and 324 male scans.
While all scans had a slice thickness of 1mm, the slice spacing varied depending on the healthcare centers protocols for storing scans.
We collected 112 scans with 1mm spacing, 84 scans with 5mm spacing, and 449 scans with 10mm spacing.
CT Scans are 3D volumes, where the $x,y,$ and $z$ axes are commonly used to refer to the anterior-posterior, lateral, and distal-proximal axes respectively. 
All studies were re-sampled prior to our collection to have an $x$ and $y$ resolution of $512 \times 512$, with varied number of slices depending on the slice spacing and the length of the distal-proximal axis.
Studies with 10mm spacing generally have 30-60 slices, where as studies with 1mm slice spacing have 150-300 slices. 

\subsection{Annotation of Pulmonary Opacification}

General and Sub-specialist Radiologists (Medical Doctors trained in medical imaging) determined the most clinically relevant classes for annotation of the data.
In order to aid prognosis and diagnosis, we segmented 6 different patterns of pulmonary opacification seen in COVID-19 pneumonia, as outlined in Table \ref{tab:classes}.
These patterns are composed of GGO, varying degrees of lobular septal thickening (an anatomic unit of the lung) and/or consolidation, and are commonly used by radiologists to differentiate between different stages and severity of infection.
Each tomographic pattern is defined and identified by its distinguishing spatial characteristics as outlined in \cite{Hansell2008}.
The aim of this project is to automate the segmentation of these 6 patterns, as the information can aid radiologists in diagnostics and evaluating patient prognostication in terms of admission to the ICU, the need for mechanical ventilation, and death.
In addition to the patterns of pulmonary opacification we also annotated pleural effusion and lymphadenopathy, both of which are non-pulmonary findings that relate to the infection. 
The annotation team was composed of a number of practicing staff and resident radiologists at Vancouver General Hospital, as well as numerous medical students at the University of British Columbia.
The medical students were used to collect lung segmentation labels, while the radiologists and residents were used to collect segmentation labels for patterns of opacification.
In order to compute the percent WAL we first segmented the total lung volumes and then the opacities within the lung. 

A team of 12 expert radiologists and residents used an online annotation tool to segment the patterns of pulmonary opacification. 
The radiologists were trained on how to use the software and instructed to annotate every slice in 10mm spaced studies, and a portion of roughly 50 representative slices with in the 5mm and 1mm spaced studies.
As a result, the thin spaced studies are partially labelled, as some slices were left unlabelled and in some regions there were no labels. 

The dataset was split on the scan level in order to ensure that all slices from a scan were contained in the same set. 
Due to the relatively few number of scans in the test set (84/663), it is composed of scans that were selected by an expert radiologist in order to ensure a clinically representative sample.
These scans were selected such that the test set included a variety of presentations, including healthy lungs, as well as lungs that contained opacities at varying stages of the disease.
In order to test the effectiveness of domain adaption, we withheld all studies from 2 locations, Italy and Vancouver, for inclusion in the test set. 
Following selection of the test set, the validation and training sets were randomly selected as 20\% and 80\% of the remaining studies, respectively.
The final number of scans in each split from each region is presented in Table \ref{tab:split}.

The labelling process for the test set differed slightly from the training set in order to increase the quality of the annotations.
The instructions for the annotators stayed the same to limit bias, however we ensured that every slice of each scan in the test set was labelled, in order compute a more accurate estimate of the ground truth percent WAL. 
In contrast to this, for the training set we prioritized getting slices labelled from more scans in order to increase the diversity of the training set, so we instructed the annotation team to label fewer slices out of each study.
This process allowed us to efficiently collect a high quality test set while balancing diversity in the training set.

\section{Annotation Data Analysis}

Standard computer vision applications and data sets, such as CamVID \cite{BrostowFC:PRL2008}, contain very little variance when annotated by different annotators. 
This is because it is rather trivial to identify simple objects such as trees, houses, etc. 
However, this is not generally the case when working with medical data sets. 
For the problem at hand, there is significant inter-observer variability, similar to other studies presented in medical imaging \cite{karimi2019deep} due to the fact that the segmentation task is non-trivial even to an expert.

\begin{figure*}[t]
    \centering
    \includegraphics[width=0.9\textwidth]{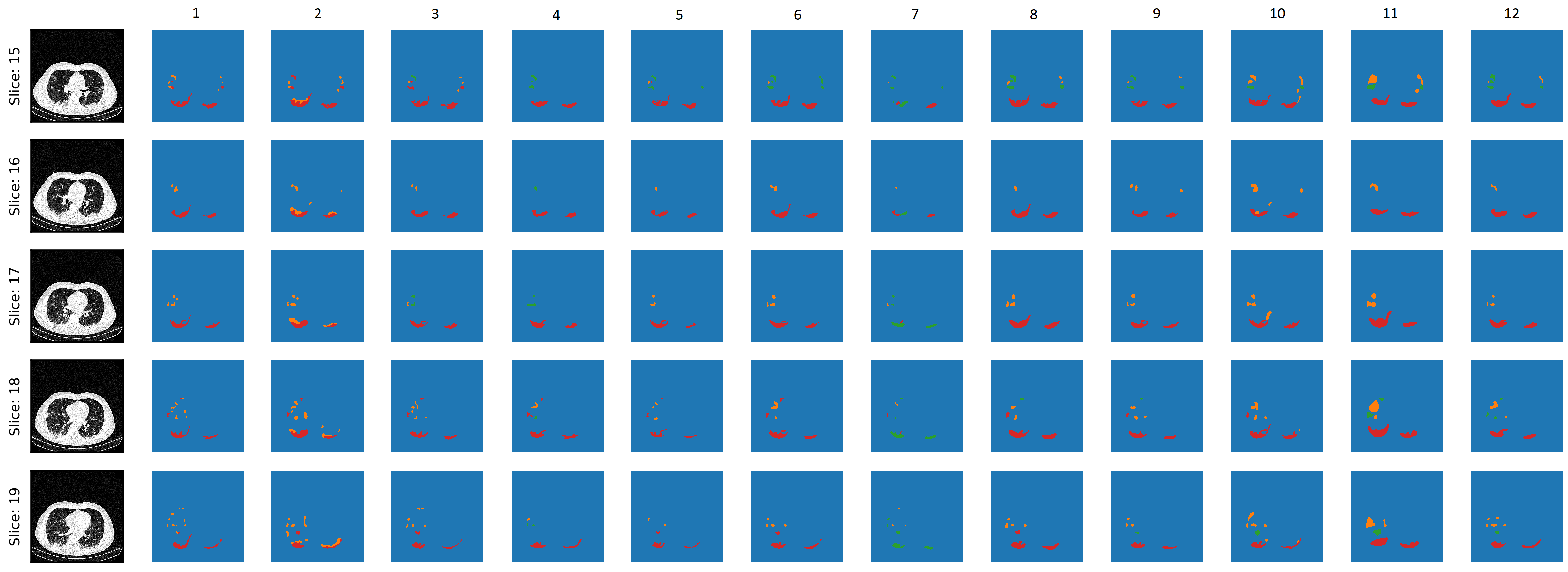}
    \caption{Comparison of the labels generated by each radiologist (columns) for 5 consecutive slices (rows) from the same study.}
    \label{fig:annotator_comparison}
\end{figure*}

\begin{figure}[t]
\centering
    \subfloat{%
        \includegraphics[width=\linewidth]{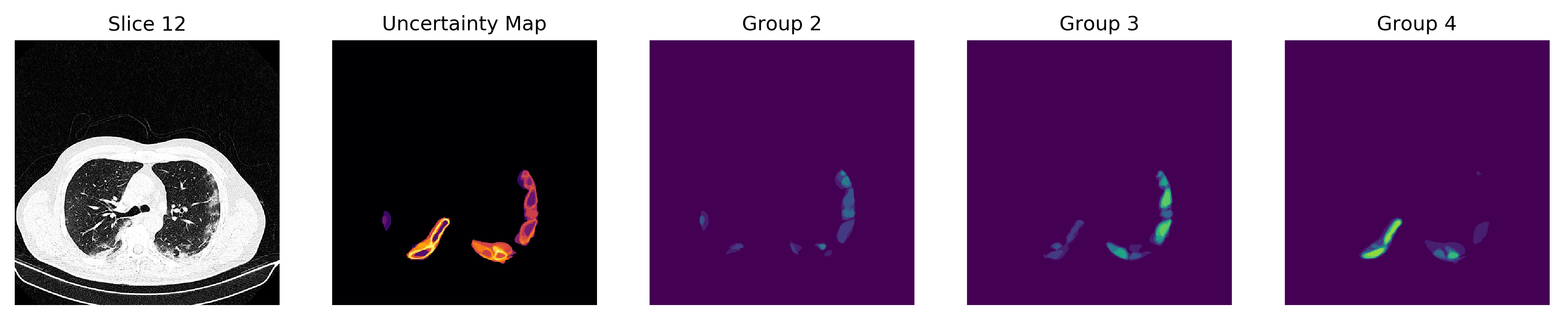}
    }
    \\
    \subfloat{%
        \includegraphics[width=\linewidth]{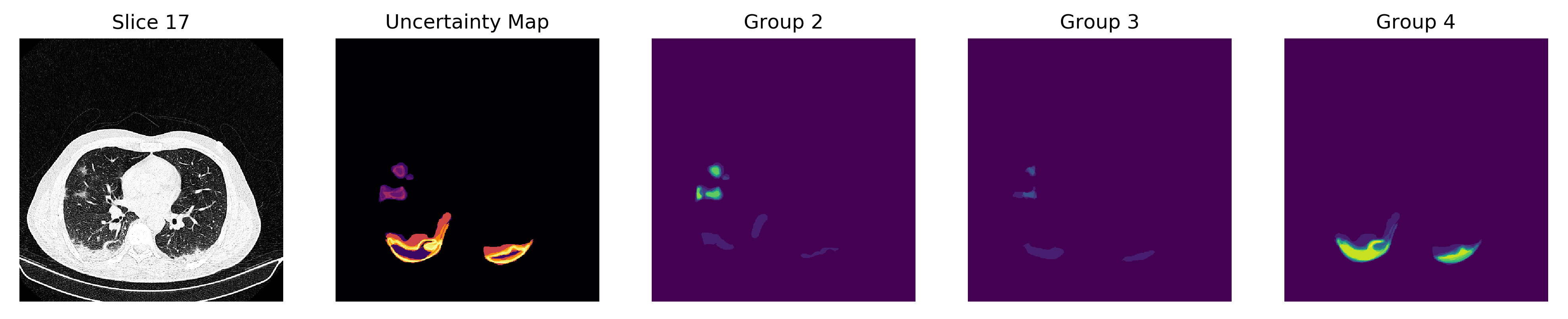}
    }
\caption{Mean and  standard deviation maps computed from  the 12 samples.}
\label{fig:probs_std}
\end{figure}

In the attempt to quantify this difference we have performed a study, collecting data from 12 different experts that annotate the same 43 slices, thus allowing us to compare and quantify the inter-observer variability present in our dataset. 
A visual comparison of the different annotations is presented in Figure~\ref{fig:annotator_comparison} and \ref{fig:probs_std}.
Qualitatively, we see that while there are differences between the radiologists they are all generally focused on the same regions of the lung.
We observe that there is large variability in both the borders of the opacity and more so in the type of opacity. 
As we discuss next, these differences require attention when training a network.


\begin{figure}[t]
\centering
    \subfloat[\label{fig:iou_matrix}]{%
        \includegraphics[width=\linewidth]{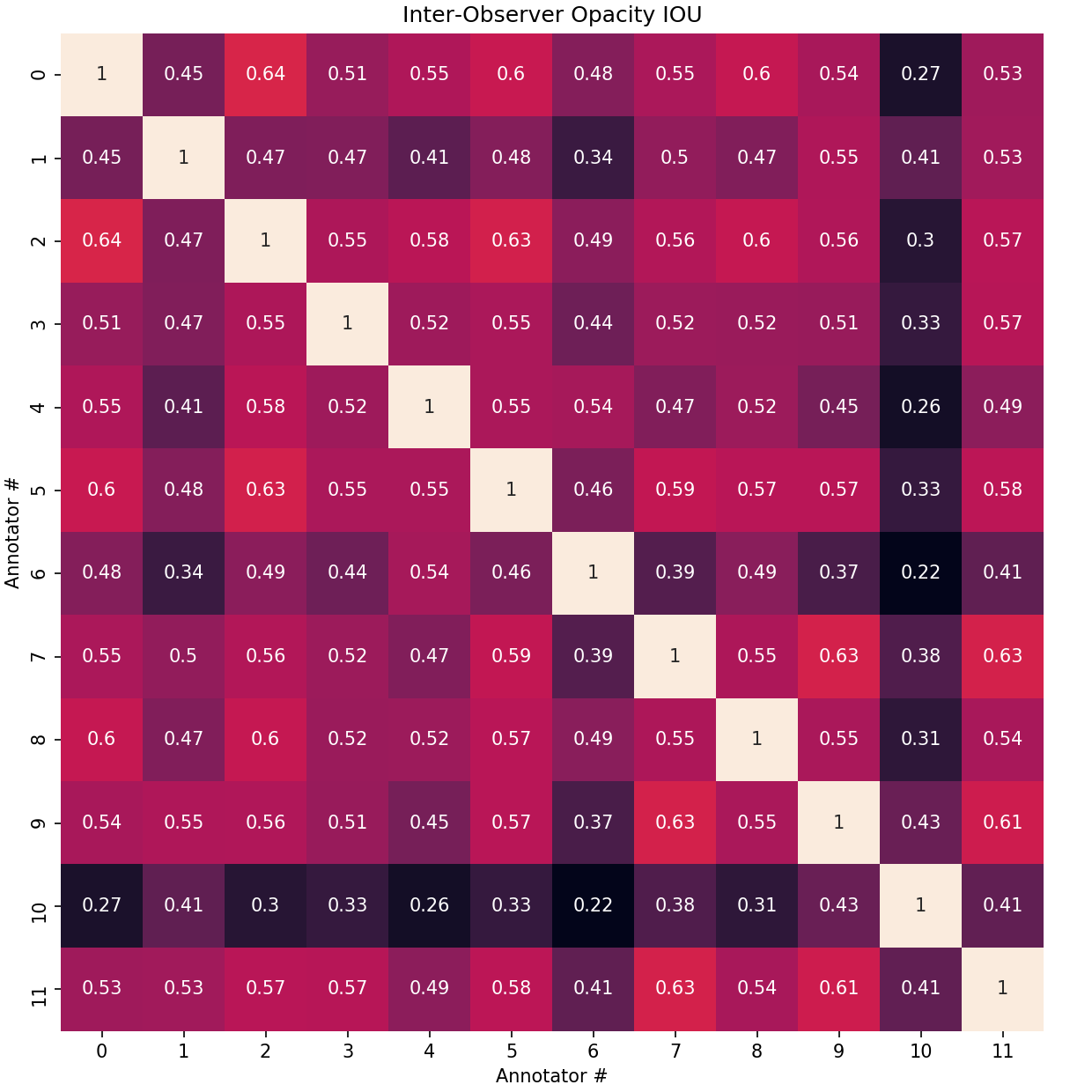}
    }
    \\
    \subfloat[\label{fig:iou_avg}]{%
        \includegraphics[width=0.9\linewidth]{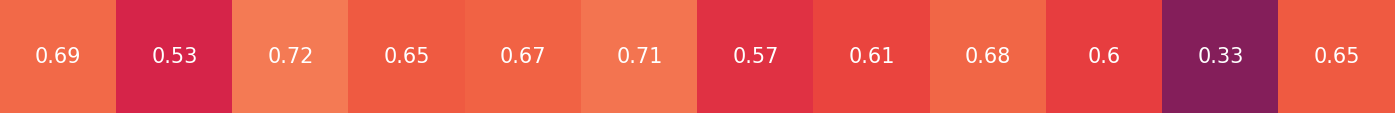}
    }
\caption{Inter-observer opacity IOU comparisons. Figure \ref{fig:iou_matrix} shows a pair-wise distance matrix between all 12 radiologists, and Figure \ref{fig:iou_avg} shows the distance between each radiologist and the average prediction. We see that, by a large margin, each radiologist is more similar to the average prediction than to any of their peers.}
\label{fig:inter_obs_iou}
\end{figure}

To see why traditional metrics may be insufficient to deal with this problem we compute the intersection over union (IOU), which is a standard metrics in semantic segmentation \cite{everingham2010pascal}. 
The results are presented in a form of a matrix in Figure~\ref{fig:inter_obs_iou}.
As can be seen in Figure \ref{fig:inter_obs_iou}, the IOU of each radiologist relative to their peers is rather poor. An AI system with similar IOU is typically considered as inoperable. 
The main reason that IOU is an inappropriate measure is that it assumes hard borders, where the problem at hand does not present such borders. 
Furthermore, since the decision on the opacification type is highly subjective, results tend to yield even lower IOU's. Thus, in the next section we discuss techniques to deal with the uncertainty in the data. 

\section{Probabilistic Labels}

By definition GGO and consolidation are hazy cloud-like opacities that do not have clear boundaries due to their physical structure, therefore they are better represented by a continuous probability distribution rather than a discrete one.
Unfortunately, commonly available annotation tools do not make it easy to incorporate uncertainty into the labelling, nor is it time efficient to ask annotators to label in such a manner.
In cases such as this, where segmentation is non-trivial and the object does not have a clear border, disagreement between hard labels should not simply be attributed to annotation error.
Instead, we view each label as a sample from the ground truth, which is an unknown underlying continuous probability distribution.
This can be modeled by 
$$ {\bf S}^{\rm obs} = \bar{\bf S} + {\bf n}({\bf x},c). $$
Here ${\bf S}^{\rm obs}$ is the observed segmentation obtained by the radiologist, $\bar{\bf S}$ is the average segmentation and ${\bf n}({\bf x},c)$ is a noise model that depends on the location, e.g. how close is the region to the edge of the object, and the class of the object, $c$.

The noise model is clearly correlated, as different classes tend to be more correlated than others as they present in a more similar way. 
It is also correlated in space, as pixels that are close to the boundary can depend on each other. 
While a comprehensive treatment of the problem is beyond the scope of this paper we have been experimenting with non-parametric noise models as well as simple parametric noise models. 
In this paper we present a simple parametric approach that uses only the first and second moments of the data estimated from the uncertainty study. 
To this end, we assume that each data point (pixel) represents a measurement from a Gaussian statistic. 
We then use the KL-divergence measure to compare the probability obtained by the network to the probability parameterized by Gaussian distribution. 
This approach takes into consideration first order statistics that is presented in the data.
We are currently collecting more data that compares different annotations of the same slice in order to develop a more comprehensive model that will allow us to train with noise models that are closer to realistic ones.

\section{Experiments}
\label{sec:experiments}

In this section we describe the various deep learning approaches we have taken to solve the problem.
Although the problem is 3D by nature, in this early phase of development we have focused on developing 2D methods that segment each axial slice independently.
While this is clearly sub-optimal, as the inclusion of the $z$ axis contains relevant information, we decided to first focus on 2D methods to set a baseline for all future approaches. 
Training 3D models is undergoing as new data arrives and results will be presented in the future.

\subsection{Data Pre-Processing}
A lung window of $-1000$Hu to $350$Hu is applied to each slice, followed by normalization of the pixel values in each slice using the mean, $-653.2$Hu, and standard deviation, $628.5$Hu, computed across the entire training set. 
In this initial work, we have grouped the 6 patterns into 3 clinically relevant grouping as outlined in Table \ref{tab:classes}. 
These groupings are created in order to increase the inter-class variation while maintaining clinical relevancy. 

\subsection{Model Architectures}
In this initial phase we focus on training three popular 2D segmentation networks, which are a natural first step for this problem and do not require any special care given the wide variety of slice spacing in the dataset.
We train a UNet \cite{ronneberger2015u}, a DeepLabv3Plus with a ResNet50 backbone \cite{chen2017rethinking},  and a PSPNet with an InceptionResNetv2 \cite{szegedy2017inception} backbone.

\subsection{Metrics}
In order to compare models we select a variety of common metrics.
Firstly, we use the  Intersection-over-Union to evaluate the accuracy of the segmentation for each class as
\begin{equation}
  IOU := \frac{TP}{TP + FN + FP}.  
\end{equation}
Although as we have previously demonstrated, this metric does not represent a desired property it is commonly used and thus we compute it to compare to known literature.

In addition, we combine the probabilities for each opacity group together to compute the \textit{Opacity IOU}, which we use to evaluate the ability of the model to distinguish between healthy lung and opacification.
In order to more sensitively determine the accuracy of the model at computing the percent WAL, we compute the ratio of the predicted volume of opacification with the ground truth, a metric we are calling the \textit{Relative Volume} (RV).
This metric is computed over the entire test set as
\begin{equation}
  RV := \frac{\bar{TP} + \bar{FP}}{\bar{TP} + \bar{FN}}.
\end{equation}
where $\bar{TP},\bar{FP}, \bar{TP}$ and $\bar{FN}$ are integrated quantities.
This metric is much less sensitive compared with the IOU as it integrates the opacities, however more sensitive than comparing the percent WAL directly because it does not depend on the lung volume, which is often much larger than volume of opacification. Therefore, unlike the IOU, even if individual boundaries do not match the overall relative volume should be correct.
Furthermore, while the IOU of an individual structure has no real clinical value, the RV has significant clinical value and therefore estimating it correctly is a much more desired goal.

\subsection{Training the Network}

In our training we use the ADAM optimizer \cite{kingma2014adam} with a batch size of $64$ to minimize the weighted KL divergence loss for $30$ epochs using a learning rate of $10^{-1}$, which is decayed by a factor of $10$ every $10$ epochs.
The loss is weighted using the complement of the probability of each class, that were computed from the training set.
The model selected for evaluation on the test set is selected using the binary opacity IOU on the validation set.

\subsection{Qualitative Results}

\begin{figure}[t]
\centering
    \subfloat[Excellent segmentation of multiple regions of pure GGO.]{%
        \includegraphics[width=0.95\linewidth]{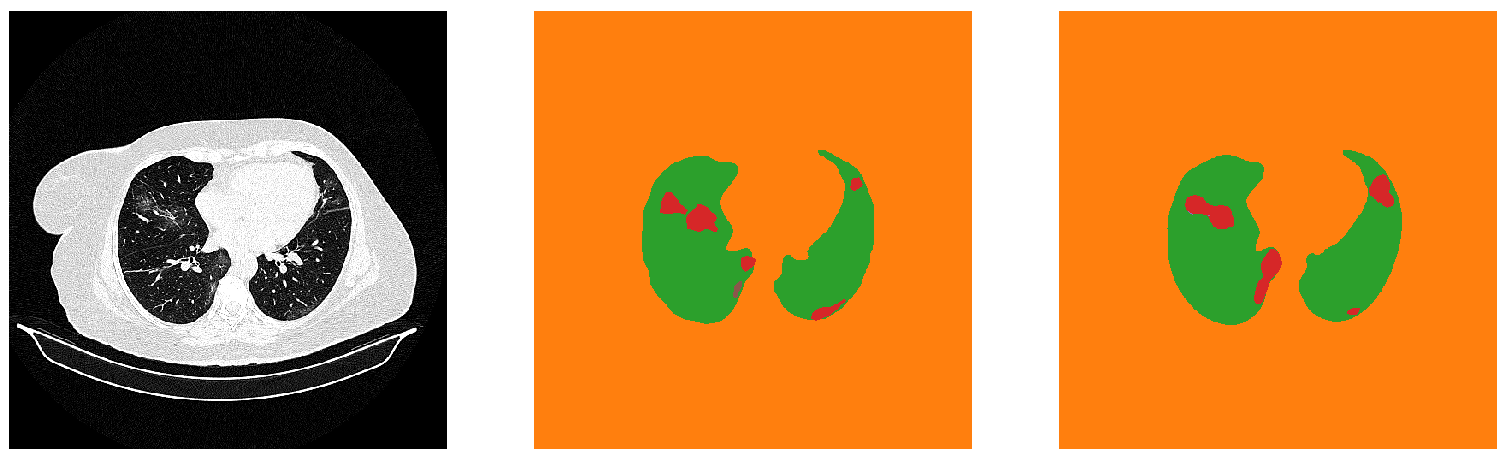}
    }
    \\
    \subfloat[Segmentation of a complex region of opacification that is composed of a mosaic of different patterns.]{%
        \includegraphics[width=0.95\linewidth]{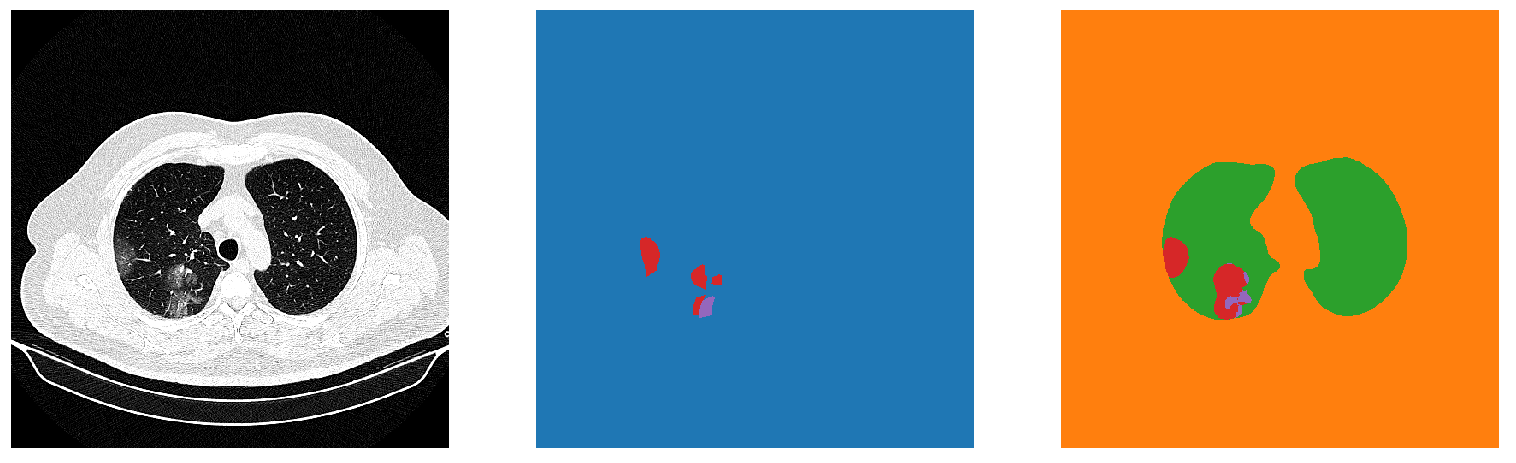}
    }
    \\
    \subfloat[An example showing how a partial volume artifact from the diaphragm causes a false positive.]{%
        \includegraphics[width=0.95\linewidth]{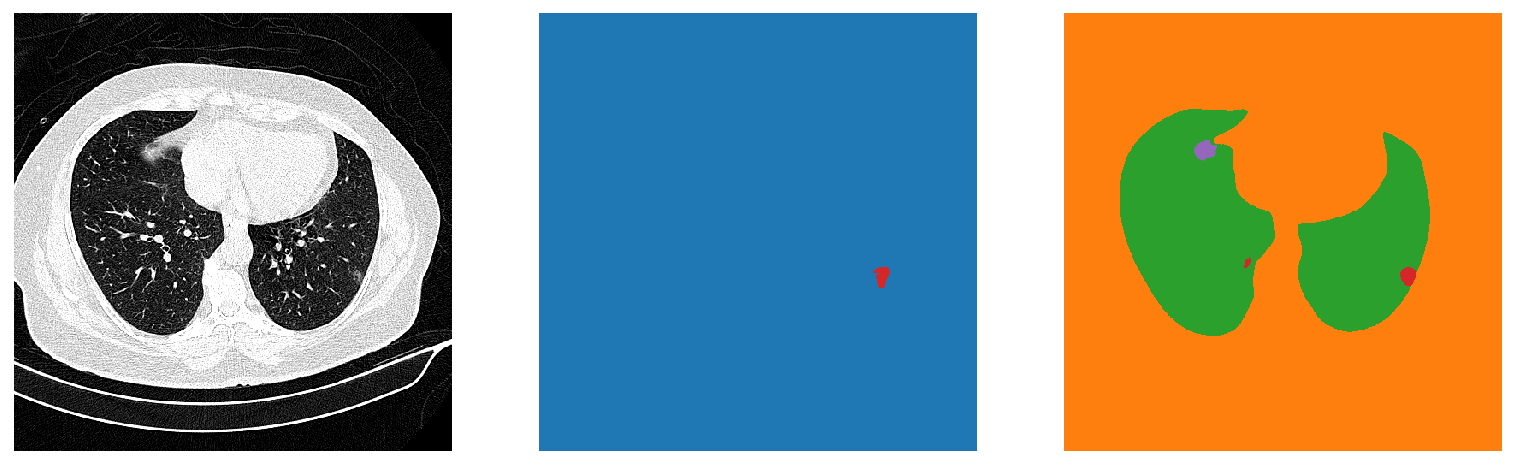}
    }
    \\
    \subfloat[In this example we see that the model is more specific than our annotator, and was able to differentiate a region of crazy paving from a larger region of consolidation..]{%
        \includegraphics[width=0.95\linewidth]{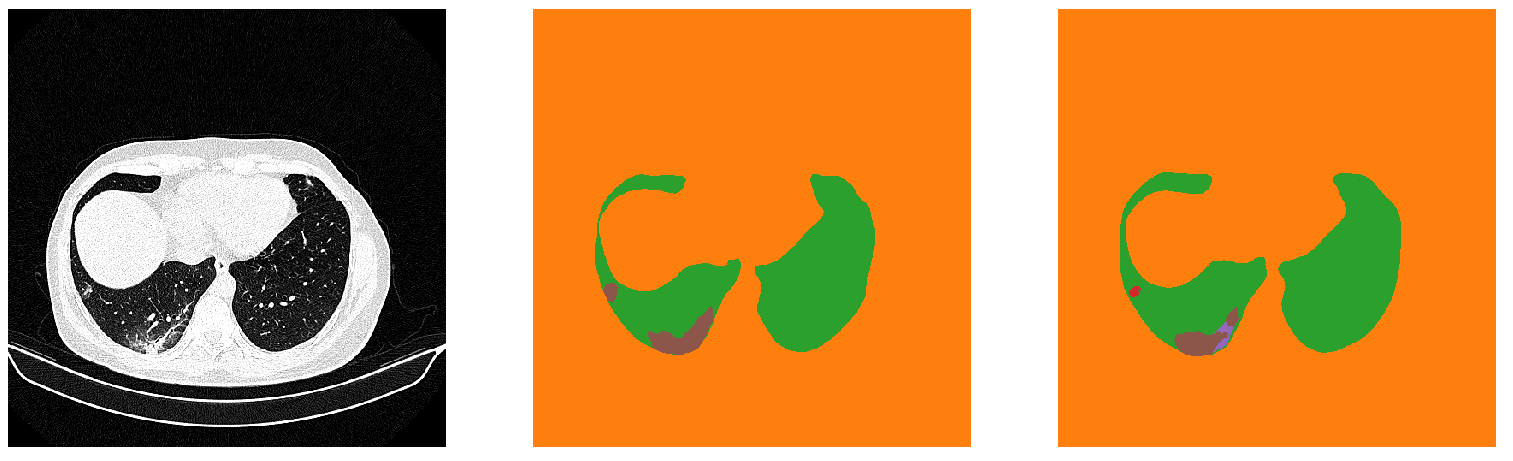}
    }
\caption{Predictions from the UNet model on four slices from the test set. The first column shows the CT scan, the second column shows the ground truth annotation, and the last column shows the model predictions.}
\label{fig:results}
\end{figure}

We visualize select output of the model in Figure \ref{fig:results}, where we see that the qualitative results are visually correct.
In the first figure we present a successful segmentation of Pure GGO.
While there are some variations in the border compared to the "ground truth" as estimated by a radiologist, all regions of opacification have been segmented by the model.
It is important to note that unlike other problems in computer vision where the ground truth is correct, it is impossible to know if the radiologist is better than our trained model given one label. 
Further studies are needed in order to obtain a full quantitative assessment of the results. 

In the second row we see a prediction on a slice, where the "ground truth" contains a combination of group 2 and group 3.
This presentation is typical where regions of GGO have progressed in severity and inter/intra-lobular lines have formed, which is a defining characteristic for Group 3.
Again, we see some disagreement between the prediction and the "ground truth" in terms of the exact border, however the model has correctly identified each region of opacification.
In qualitative terms, the prediction is equally viable to the ground truth and the difference between our segmentation to the "ground truth" is similar to the one obtained by the segmentation by different radiologists.

In the third row we show the effects of partial volume artifacts, which we consider to be the most common cause of false positive. 
In this case we see the effect of the diaphragm causing a opaque region in the anterior portion of the right lung, however it is also common to see partial volume artifacts at the apex of the lungs as well.
Seeing as a radiologist would rule this out by viewing adjacent slices and recognizing the start of the diaphragm, a potential solution for this would be to include information from additional slices, such seen in 3D approaches.

In the bottom row we see the successful segmentation of consolidation in the posterior basal segment of the right lower lung lobe.
We would like to highlight the region of the opacification that has been predicted to be group 3 (purple), however was labelled as group 4 (brown).
In consultation with a second radiologist we have confirmed that the prediction can be seen as more correct than the label in the predicted data set.
The likely cause of this was previously discussed, where we show that more specific options are needed by the annotators.
In this case since the region of opacification was mainly group 4, it is likely that the annotator simply labelled the entire region as such.

\subsection{Quantitative Results}
\begin{table*}[t]
\small
\centering
\caption{Quantitative Segmentation Results on the Test Set}
\begin{tabular}{lrrrrrr}
\hline
\textbf{Model} &
  \multicolumn{1}{l}{\textbf{Loss}} &
  \multicolumn{1}{l}{\textbf{Relative Volume}} &
  \multicolumn{1}{l}{\textbf{Opacity IOU}} &
  \multicolumn{1}{l}{\textbf{Group 2 IOU}} &
  \multicolumn{1}{l}{\textbf{Group 3 IOU}} &
  \multicolumn{1}{l}{\textbf{Group 4 IOU}} \\ \hline
DeepLabv3Plus (ResNet50)   & 0.396 & 0.967 & 0.706 & \textbf{0.423} & 0.375 & 0.428 \\
PSPNet (InceptionResNetv2) & \textbf{0.377} & 1.044 & 0.735 & 0.414 & \textbf{0.390} & 0.391 \\
UNet                       & 0.389 & \textbf{1.017} & \textbf{0.758} & 0.370 & 0.317 & \textbf{0.432}
\end{tabular}
\label{tab:results}
\end{table*}

In order to compare the different deep learning models, quantitative results are presented in Table \ref{tab:results}.
The UNet proved to be the best segmentation model in terms of opacity IOU, while we see that the other models performed similarly, yet slightly worse.
No single model stands out as having the best mIOU over the three opacity groups, and all scores are relatively low. 
Given the relatively high opacity IOU, this could be attributed to a significant confusion between opacity groups.
Despite varied IOU scores, the models all predicted clinically relevant relative volumes of opacity, with the UNet and PSPNet over predicting by 1.7\% and 4.4\% respectively, and the DeepLabv3Plus under predicting by 3.3\%. 

\section{Discussion}
\label{sec:discussion}

Despite the challenging task of segmenting pulmonary opacification on CT scans, these initial promising results already indicate that deep learning approaches can provide value in clinical situations.
All models were able to achieve relative volumes ratios within $\pm 5\%$ of the ground truth, enabling clinically relevant automatic estimation of the percent WAL.
We show qualitative results that our model is able to determine the pattern of opacification, which could provide timely and valuable clinical information to healthcare centers, as these patterns are associated with the severity of the infection.
We do not see large differences between different model architectures observed in \cite{fan2020inf}, suggesting that the quantity and diversity of data are more important that the network architecture, as proposed in \cite{hofmanninger2020automatic}.
Success on the test set, which contains scans from regions that were held out of the training set, shows that the models are able to successfully transfer knowledge across domains.
This is an important finding, as the sharing of pre-trained models may allow the community to circumvent the lack of publicly available data due to privacy restrictions.

The best model achieves an opacity IOU of 0.758 on the test set, which we note is slightly higher than the human level performance we found comparing each radiologist to the average prediction in Section \ref{sec:dataset}.
Given the analysis of the inter-observer variability, we believe these models have approached the upper bound of the accuracy for our labels, without including higher order statistics or better data.
Thus, given a test set of similarly collected labels we are not able to reliably evaluate quantitative performance using simple statistics beyond this point.
we see that comparison of qualitatively ideal annotations from expert annotators yield a wide range of IOU scores.
While it is noteworthy that the model's opacity IOU is higher than the values we see between radiologists and the average label, due to the fact that the human level analysis comes from only one scan we do not believe it is possible to determine if the model has surpassed human ability yet.

The relatively poor quantitative performance in differentiating between pattern of opacification is likely due to a number of factors.
Firstly, in Section \ref{sec:dataset} we see large disagreements between experts on the specific pattern of opacification in a given region, as shown in Figure \ref{fig:probs_std}.
While it is possible that annotator error is a factor, we believe that this could largely be rectified through improved annotation procedure.
Internal patterns within the segmented regions can mix together like a mosaic of many patterns in a single segmented region, and therefore should not be characterized by a single number. 
A more appropriate description should be given by a partial volume or probability, however creating such labels is difficult with software tools commonly used for annotation. 
Thus, currently, our labels do not always account for the variations in the type of pattern across regions making the labels inherently noisy, which effects training and evaluation of the model.
In future work we plan to adjust the labelling procedure to provide annotators with more specific tools and instructions, such that we are able to obtain more accurate labels.
Lastly, even with the groupings described in Table \ref{tab:classes}, there is low inter-class variability between the patterns of opacification, meaning that even given excellent annotations the segmentation task is likely challenging.

This application demonstrates the need to further develop techniques that allow the use of noisy labels with a non-trivial noise model. A simple example is the use of covariance information that is non-trivial to incorporate when using Stochastic Gradient Descent. 

\section{Conclusion}
\label{sec:conclusion}
In this paper we have described our collection and annotation of CT scans of patients with COVID-19 pneumonia for the segmentation of patterns of pulmonary opacification.
We provide results using three popular 2D segmentation networks that show that we are able to accurately compute the relative volume of opacity, and estimate the composition of three clinically relevant pattern groups that have been linked to patient outcome.
We show that our models are able to adapt to new domains by withholding two regions from training, which provides valuable insight on the possibility of a community driven approach that is not hindered by the lack of publicly available data or restrictive privacy policies.
We provide an analysis of the inter-observer variability that is present in this non-trivial task, and conclude that due to the physical structure of opacification and the low inter-class variability, improved success in this task requires the adoption of soft labelling techniques and probabilistic models.
To this end, we propose improved annotation procedures and noise modelling techniques that allow for future work using continuous probability distribution as opposed to the more common discrete distributions used in computer vision. 

\section*{Acknowledgment}
We thank Brian Lee and Duncan Ferguson for their support in collecting and creating the dataset, Vancouver Coastal Health Research Institute and the doctors, medical students, and healthcare centers who contributed to the creation of the dataset. For a comprehensive list please see \href{https://cic.ubc.ca/projects/open-source-ai-model-for-covid-19-ct-diagnostics-and-prognosis/acknowledgements/}{our acknowledgments page}. 
K.L and E.H. are supported of the Natural Sciences and Engineering Research Council of Canada (NSERC). 

{\small
\bibliographystyle{abbrvnat}
\bibliography{refs}
}

\end{document}